\documentstyle[12pt]{article}
\begin{document}

\title{Transport properties of one-dimensional Kronig-Penney models with correlated
disorder}
\author{Tsampikos Kottos$^1$,G. P. Tsironis$^1$ and Felix M. Izrailev$^{1,2}$\thanks{%
email addresses: izrailev@vxinpb.inp.nsk.su ; izrailev@physics.spa.umn.edu} 
\\
\\
$^1$ Department of Physics, University of Crete\\
and Research Center of Crete, P.O. Box 1527\\
71110 Heraklion-Crete, Greece \\
\\
$^2$ Budker Institute of Nuclear Physics,Novosibirsk 630090, Russia}
\maketitle

\begin{abstract}
Transport properties of one-dimensional Kronig-Penney models with binary
correlated disorder are analyzed using an approach based on classical
Hamiltonian maps. In this method, extended states correspond to bound
trajectories in the phase space of a parametrically excited linear
oscillator, while the on site-potential of the original model is transformed
to an external force. We show that in this representation the two probe
conductance takes a simple geometrical form in terms of evolution areas in
phase-space. We also analyze the case of a general $N$-mer model.
\end{abstract}


\baselineskip 24 pt \vspace{2in} \newpage

\newpage

\baselineskip 24 pt \vspace{2in} 
\newpage

\newpage

\section{\bf Introduction}

The random dimer model, a tight binding model with correlated disorder,
introduced in refs. \cite{Dunlap,Phillips} has attracted considerable
attention due to the presence of transparent states in an otherwise
disordered one dimensional system \cite{Dunlap}-\cite{IKT95}. In the present
paper we address the issue of the spectrum of the random dimer Kronig-Penney
(RDKP) model \cite{SMA94} and extensions using the same Hamiltonian approach
applied earlier in the context of the tight binding models \cite{IKT95}.
Using techniques from dynamical systems theory \cite{BFLT82}, we construct a
Poincare map that turns the Kronig-Penney model into an equivalent
tight-binding model and study the latter through a two-dimensional map
corresponding to a classical linear oscillator with a parametric
perturbation given in the form of periodic delta-kicks \cite{IKT95}. The
amplitudes of these kicks are defined by the site potential of the
tight-binding model. In this representation, extended states of the
tight-binding model are represented through bounded trajectories in the
phase space of the Hamiltonian map. Furthermore, in this representation the
two probe conductance is related to the time evolution (under the
Hamiltonian map) of areas initially defined by the basis unit vectors. This
new approach provides an effective and simple tool for understanding
transport properties, the structure of eigenstates as well as for deriving
analytical expressions. In particular, one can easily determine fully
transparent states for the general case when $N$ -sites are correlated. In
the following section we summarize briefly the Hamiltonian map approach used
in ref. \cite{IKT95} and apply it in the context of the RDKP model. In
section 3 we analyze transport properties through a new expression for
conductance while in section 4 we conclude.


\section{\bf Hamiltonian Map Approach}


\subsection{\bf Time-Dependent Linear Map}

The model of interest is the one-dimensional Schr\"odinger equation with an
array of $\delta-$function potentials: 
\begin{equation}  \label{tbe1}
E \phi (z) = -\frac {d^2 \phi (z)}{d^2z} + \sum_{n=1}^L \epsilon_n \delta
(z-z_n) \phi (z).
\end{equation}
Equation~(\ref{tbe1}) defines the Kronig-Penney model where $E$ is the
eigenenergy of the stationary states, $\epsilon_n$ is the strength of the
potential and $z$ denotes space while we take the positions of the $\delta$
functions to be regularly spaced $(z_n = n)$. The tight-binding model
corresponding to Eq.~(\ref{tbe1}) is \cite{BFLT82} 
\begin{equation}  \label{tbe2}
\phi_{n+1}+\phi_{n-1}=v_n \phi_n , \,\,\,\,\,\,\, v_n=2 cos(q) + \epsilon_n {%
\frac {sin(q)}{q}}
\end{equation}
with $q^2 \equiv E$, $\phi_n \equiv \phi(z=n)$; Eq. (\ref{tbe2}) can be
written equivalently as a two-dimensional map $M_n$, i.e.: 
\begin{equation}  \label{eq:mmap}
\left( 
\begin{array}{c}
x_{n+1} \\ 
y_{n+1}
\end{array}
\right) =\left( 
\begin{array}{cc}
v_n & -1 \\ 
1 & 0
\end{array}
\right) \left( 
\begin{array}{c}
x_n \\ 
y_n
\end{array}
\right) \,,
\end{equation}
where $\phi_n = x_n$ and $y_n=\phi_{n-1}$. An eigenstate of Eq. (\ref{tbe2})
is a 'trajectory' of the map of Eq. (\ref{eq:mmap}). Straightforward
diagonalization of this map leads to the eigenvalues ${\lambda }_n^{\pm }=%
\frac{v_n\pm i\sqrt{4-v{_n}^2}}2=e^{\pm i\mu _n}$ where the phase $\mu _n$
is introduced by the relation $v_n=2\,cos\mu _n$. For $|v_n|<2$ we obtain
stable map rotation with phase $\mu _n$. For $|v_n| =2$ we find the curves
separating regions of allowed and forbidden energies, determined through
equations $q = (2k+1) \pi$, for $\epsilon = -2 q cot(q/2)$ and $q = 2k \pi$,
for $\epsilon = 2 q tan(q/2)$ with $k=0,1,2,\ldots$. To understand the
origin of the resonant states in the RDKP model resulting when adjacent
pairs of random energies $\epsilon_n$ coincide \cite{SMA94}, we consider the
sequence ${\epsilon _n}$ which consists of one dimer only i.e. where all the
values of $\epsilon _n$ are equal to $\epsilon _1$ except two values for
which we have $\epsilon _m=\epsilon _{m+1} =\epsilon _2$. From the property
of the map eigenvalues found previously we observe that this unique dimer
with energy $\epsilon _2$ does not influence the trajectories of the map of
Eq. (\ref{eq:mmap}) when the total phase advance $\mu _m+\mu _{m+1}=2\mu _m$
through the dimer is equal to $\pi $ or 2$\pi $ . Since the latter value $%
\mu _m=\pi $ is forbidden from the stability conditions, the resonant energy 
$q_{cr}^2 \equiv E_{cr}$ is defined by $\mu _m=\pi /2$ giving $-2
q_{cr}/\epsilon _2 = tan (q_{cr})$. As a result, for the general case of
randomly distributed dimers $\epsilon _1$ and $\epsilon _2$, there are two
resonant values, 
\begin{equation}  \label{dimer}
q_{cr}=-{\frac {\epsilon _{1,2} tan (q_{cr})}2}
\end{equation}
for which dimers of the first type, $\epsilon _1$ , or of the second, $%
\epsilon _2$ , have no influence on the transparent states. One should note
that since $tan(q_{cr})$ is a $\pi -$periodic function and takes values in $%
(-\infty,+\infty)$, we have an infinite set of critical energies $q_{cr}$
(two in every interval $[(2k-1)\pi/2, (2k+1)\pi/2]$) \cite{SMA94}. However,
for the first allowed band defined through the condition $|v_n|=2$, the
disorder strength $\epsilon$ must be greater than a critical value $%
\epsilon_{cr} = -2$ in order to have resonant states. The critical disorder
for reflectionless modes to appear, arises only in the first band of the
spectrum.

The previous analysis can be extended to the general case of $N$-mer (two
values $\epsilon _1$ and $\epsilon _2$ appear in blocks of length $N$) where
the resonant energy is defined through the following condition 
\begin{equation}  \label{general}
\mu_N =\frac \pi N,\frac{2\pi }N,\frac{3\pi }N,\frac{4\pi } N,\cdots ,\frac{%
(j+1)\pi }N;\,\,\,\,\,\,\,\,\,\,\,j=0,1,2,....N-2.
\end{equation}
We note that as the block size $N$ increases the number of resonant states
proportional to $2(N-1)$ increases as well. In the first zone, in
particular, the disorder strength should be smaller than a critical strength $%
\epsilon_{cr}(N)$ for this to happen. The latter is obtained through
equation $2 cos(q_{cr})+ \epsilon _{1,2} \frac {sin(q_{cr})} {q_{cr}} =
2\cos \mu _N$ that gives the corresponding critical wavevectors.
Equivalently, $\epsilon_{cr}$ is obtained through: 
\begin{equation}  \label{phase2}
\frac {sin q}{cos q - cos \mu_N} = -\frac {2 q} {\epsilon_{cr}}.
\end{equation}
The resonances inside the first band appear whenever the derivative of the
left hand side of Eq.(\ref{phase2}) at $q=0$ is less than $-\frac {2}
{\epsilon_{cr}}$ and thus 
\begin{equation}  \label{phase3}
\epsilon_{cr}(N) = 2 (cos(\mu_N) - 1)
\end{equation}
In the phase diagram of Fig.~1 we distinguish three regions: (I) contains no
resonant states, (II) where some resonant states appear and (III) where all
resonants are concentrated. The borders of region (II) start at $\epsilon =
-2$ for $N = -2$ (RDKP case) and are given by the curve Eq.~(\ref{phase3})
with $\mu_N=\pi/N$ (upper bound) and $\mu_N=(N-1)\pi/N$ (lower bound) and
approach zero and $-4$ respectively as $N \rightarrow \infty$. We recall
that for the perfect Kronig-Penney lattice the range of accessible $\epsilon$
values within the first energy band is $[-4, +\infty )$. We further note
that the lower border corresponds to the existence of only one resonant
state, while as we are going towards the upper border more resonances appear.
The upper critical curve delimits region III in which all system resonances 
are found.

By introducing a new variable $p_{n+1}=x_{n+1}-x_n$ playing a role
similar to momentum we obtain through Eq.~(\ref{eq:mmap}) a new map 
representation: 
\begin{equation}  \label{mmap1}
\begin{array}{cc}
p_{n+1}=p_n+f_n\,x_n &  \\ 
x_{n+1}=x_n+p_{n+1} & 
\end{array}
\end{equation}
where $f_n=v _n - 2$ and having the same eigenvalues with the original map (%
\ref{eq:mmap}). In the map of Eq.~(\ref{mmap1}) ellipses correspond to $%
\epsilon_n=\epsilon_1$ for all $n$, a defect at site $m$ with $\epsilon_m=
\epsilon_2$ results in a kick into another ellipse (Fig.~2a) while correlated
defects lead to a return to the original ellipse, since the total phase 
advance $\mu = \mu_m +\mu_{m+1} = \pi$  (Fig.~2b). When a random mixture of 
dimers with
energy $\epsilon _2$ is embedded in a chain with energy $\epsilon _1$, we
obtain a phase space trajectory similar to the one in Fig. 2c. We observe two 
ellipses corresponding to $\epsilon _1$ and $\epsilon _2$ values respectively.
The second ellipse is formed by points occurring every time the first site
of a dimer is encountered by the map. In the general case of correlated $M-$%
block under the condition that the total sum $\sum_{n=1}^M \mu_n$ of phase
shifts is equal to $m \pi $, for any sequence with $m$ integer, the
trajectory always returns to the ellipse associated with the "perfect" sites
to the left and to the right of the scattering potential.

\subsection{\bf Parametric Linear Oscillator}

Another useful representation of the original model of Eq. (\ref{tbe2})
similar to the map of Eq.~(\ref{mmap1}) but more convenient for the analysis
of the localization length, can be obtained through two successive maps \cite
{IKT95}: 
\begin{equation}
\begin{array}{cc}
\tilde{p}_n=p_n\,+A_n\,x_n &  \\ 
\tilde{x}_n=x_n & 
\end{array}
\label{mmap2a}
\end{equation}
and 
\begin{equation}
\begin{array}{cc}
p_{n+1}=\tilde{p}_n\cos \mu _0\,\,-\,\,\tilde{x}_n\sin \mu _0 &  \\ 
x_{n+1}=\tilde{p}_n\sin \mu _0\,\,+\,\,\tilde{x}_n\cos \mu _0 & 
\end{array}
\label{mmap2b}
\end{equation}
When the maps of Eq.~(\ref{mmap2a}) and Eq.~(\ref{mmap2b}) are combined,
they result in a form of Eq.~ (\ref{tbe2}), viz.\thinspace 
\begin{equation}
\label{mmap2}
 x_{n+1}+x_{n-1}=(2\cos \mu _0\,+\,A_n\sin \mu _0)\,x_n . 
\end{equation}

Comparing with Eq.~(\ref{tbe2}) one can establish the correspondence 
$\mu _0=q;\,\,\,\,\,\,\,\,\,\,\,\,\,\,A_n=\frac{\epsilon _n}q $ 
between the parameters $q,\epsilon _n$ in the original model (\ref{tbe2})
and the parameters $\mu _0,A_n$ of the map of Eq.~(\ref{mmap2a},\ref{mmap2b}%
). The latter map has a clear meaning since the map of Eq.~(\ref{mmap2a})
corresponds to an instant linear kick of the strength $A_n$ resulting in the
change of the momentum $p_n$ and the map of Eq.~(\ref{mmap2b}) describes
free rotation in the phase plane $(p,x)$ defined by the angle $\mu _0$. The
dynamical system modeled by Eqs.~(\ref{mmap2a}),(\ref{mmap2b}) is that of a
linear oscillator with a periodic parametrical perturbation with a
Hamiltonian \cite{IKT95} 
\begin{equation}
\tilde{H}=\frac{\mu _0p^2}2\,+\,\frac{\mu _0x^2}2\,-\frac 12\,x^2\tilde{%
\delta}_1(t)\,;\,\,\,\,\,\,\,\,\,\tilde{\delta}_1(t)\equiv \sum_{n=-\infty
}^\infty A_n\delta (t-n)  \label{ham2}
\end{equation}
We note that by integrating Eq.~(\ref{tbe1}) between two successive $\delta $%
-kicks of the potential, we obtain Eq.~(\ref{mmap2b}), while integration
over a kick, results in Eq.~(\ref{mmap2a}) with a new kick strength $%
A_n=\epsilon _n$. By comparing Eq.~(\ref{tbe1}) and Eqs~(\ref{mmap2a}),(\ref
{mmap2b}) we find the significance of the variable $p_n$: it is the rescaled
(with respect to $q=\sqrt{E}$) first derivative of the local amplitude
function $\phi $ just before the $n-$th kick i.e. $p_n=(d\phi /dz)_{z=z_n}/q$.

For the dimer case defined by the two values of $\epsilon $$_1,\epsilon _2,$
we can set without loss of generality $\epsilon $$_1=0$. As a result, the
motion corresponding to $\epsilon $$_n=\epsilon _1$ is represented by the
circle in the phase plane $(p,x)$ and resonant behavior results when, after
a given number of kicks with $\epsilon $$_n=\epsilon _2,$ the trajectory
returns  to this circle. An example of this behaviour is given in Fig. 3a
for $q=q_{cr}=11.4$ and $\epsilon _2=9.75885$. A similar behaviour for the
case of a trimer $N=3$ is illustrated in Fig. 3b for $\epsilon _2=1.25$ and $%
q=q_{cr}=1.489$.  We observe that the trajectory is bounded in the
phase-space.

The above scheme is also valid for the much more general case where the
locations of the $\delta -$ functions $z_n$ in the original model Eq.~(\ref
{tbe1}) are not equidistant and can be taken from an arbitrary distribution.
In this case Eq.~(\ref{mmap2b}) has the same form provided we make the
substitution $\mu _0\rightarrow \mu _{T_n}=qT_n$, where $T_n$ corresponds to
random periods of the kicks in the Hamiltonian approach (\ref{ham2}). In
the original model of Eq. (\ref{tbe1}), $T_n$ indicates the random distance
between two successive lattice sites, i.e. $T_n=z_{n+1}-z_n$. The critical
value $q_{cr}$ for the dimer is obtained directly from Eq.~(\ref{mmap2})
using the relation $2\cos \mu _N=2\cos \mu _{T_n}+A_n\sin \mu _{T_n}$ with $%
\mu _N=\pi /2$ (see Eq.~(\ref{general})). As a result, we obtain $\tan
(q_{cr}T_n)=-\epsilon /(2q_{cr})$ where both $T_n$ and $\epsilon _n$ are
related to the same lattice site $z_n$ in Eq.~(\ref{tbe1}). Therefore, we
conclude that in the case of a generalized dimer where time displacements $T_n
$ and on-site potential are paired, in a way such that $T_n=T_{n+1}=T_2$ and $%
\epsilon _n=\epsilon _{n+1}=\epsilon _2$, the condition for the critical
energy will be similar to the one obtained previously but with the change 
$\mu _0\rightarrow \mu _{T_n}=\mu _2=qT_2.$

\subsection{\bf Nearly Resonant States}

The representation of the model of Eq. (\ref{tbe2}) used in section 2.2 allows
for the study of global properties of eigenstates. In particular, the
resonant delocalized states correspond to a bounded motion described by the
maps of Eqs (\ref{mmap1}),(\ref{mmap2a}),(\ref{mmap2b}). Localized states on
the other hand are represented by unbounded trajectories that escape from
the origin of phase space $(p,x)$. This is illustrated in Fig. (4) for the
case of random dimers with non-resonant values of $q$ . The exponential
increase of a distance from the origin $(p=x=0)$ is related to the
localization length of the corresponding eigenstate. In order to study the
dependence of the localization length $l$ for nearly-resonant states, it is
useful to pass to action-angle variables $(r,\theta )$ for the map of Eq. ((%
\ref{mmap2a}, (\ref{mmap2b}) using the definitions $x=r\,\cos \theta \,$ and$%
\,\,\,\,p=r\,\sin \theta $. We obtain a map for the action $r$ given by 
\begin{equation}
r_{n+1}^2=r_n^2D_n^2;\smallskip\ D_n^2=(1+A_n^2\cos {}^2\theta _n+A_n\sin
2\theta _n)  \label{rmap}
\end{equation}
where the transformation for $\cos \theta _n$ and $\sin \theta _n$ is given
by the relations 
\begin{equation}
\begin{array}{cc}
\cos \theta _{n+1}=D_n^{-1}\{\cos (\theta _n+\mu _0)-A_n\cos \theta _n\sin
\mu _0\} &  \\ 
\sin \theta _{/n+1}\,=D_n^{-1}\{\sin (\theta _n+\mu _0)\,+A_n\cos \theta
_n\cos \mu _0\} & 
\begin{array}{c}
\end{array}
\end{array}
\label{cossin}
\end{equation}
The relations of Eq. (\ref{rmap},\ref{cossin}) can be used instead of the
common transfer matrix approach for the determination of the localization
length $l$ . The latter is equal to the inverse of the Lyapunov exponent $%
\gamma $ defined as 
\begin{equation}
\gamma =\lim_{N\rightarrow \infty }\frac 1N\sum_{n=0}^{N-1}\ln \,(\frac{%
r_{n+1}}{r_n})  \label{gamma}
\end{equation}
where the ratio $r_{n+1}/r_n=D_n$ is given by (\ref{rmap}).

The advantage of this approach in the finding of the Lyapunov exponent $%
\gamma $ in comparison to the standard transfer matrix method is that there
is no divergence during iterations. It is interesting to note that Eqs. (\ref
{rmap}), (\ref{cossin}) can be mapped into a one-dimensional map $\theta
_{n+1}=F(\theta _n)$ which is non-linear for the non-zero perturbation $%
A_n\neq 0$. One can show that in such a representation, the expression (\ref
{rmap}) is directly related to the stretching of the phase, $d\theta
_{n+1}/d\theta _n=D_n^2$. Therefore, the original quantum problem is reduced
to the study of the properties of a one-dimensional time-dependent map and
its tangent space.

Due to correlations in the sequence $\theta _n$, the expression (\ref{gamma}%
) can not be evaluated directly. However, it is possible to construct an
effective map for two successive kicks of the single map (\ref{rmap}) and
neglect the correlations between the phases $\theta _{n+2}$ and $\theta _n$
near the resonance $q=q_{cr}-\delta \approx q_{cr}$. Applying the resulting
two-step map \cite{IKT95} to the present case allows us to estimate the
Lyapunov exponent for $\delta \ll 1$, using the expansion in $W=A_n\delta
/\sin \mu _0,$ with the successive averaging over $\theta _n,$ leading to 
\begin{equation}
\gamma \approx Q\frac{\delta ^2\epsilon _2^2}{\mu _0^2sin^2\mu _0}
\label{estimate}
\end{equation}
The factor $Q$ stands for the probability for the dimer of the second kind
(with energy $\epsilon _2$), to appear. In Fig. 5 we compare the analytical
result (\ref{estimate}) (solid line) with the numerical data obtained from
the map (\ref{rmap},\ref{cossin}) (circles) after iterating up to 4000000
time steps and averaging over more than 1000 realizations of the random kick
strengths $A_n$, for the case $Q=0.5,\epsilon =9.75885$ and $q_{cr}=11.4$.
The agreement between theory and numerical data is extremely good as seen in
Fig. 5 verifying that expression (\ref{estimate}) valid for $\delta \ll 1$.

From Eq.~(\ref{estimate}) we determine the dependence of the inverse of the
localization length for the near-resonant states and the way it changes when
the system parameters change. We note that the higher the order of the
resonance, $q>>1$ (i.e. the higher $k$ in $[(2k-1)\pi /2,(2k+1)\pi /2]$),
the larger is the localization length, and thus better transport properties are 
expected. Such a behaviour of $l(q)$ is expected since for $q>>1$ the
second term of $v_n$ (where $q$ appears in the denominator) in Eq.~(\ref
{tbe2}) becomes negligible resulting to a tight-binding equation with zero
on-site potential. The localization length increases also when decreasing $%
\epsilon $ towards zero. This is easily comprehended since when $\epsilon
_1=\epsilon _2=0$ we recover the properties of the perfect lattice. Finally,
as the concentration $Q$ of dimers decreases, the value of the
localization length for the near resonant states increases.


\section{\bf Transport Properties.}


In this section we examine the transport properties of our system by
studying the behavior of the transmission coefficient. We assume that the
system of Eq.~(\ref{tbe2}) is a sample consisting of $L$ lattice points with
two identical semi-infinite perfect leads on either side. As a result, the
left lead extends from $-\infty <n\leq 0$, the sample extends from $1\leq
n\leq L$ and the right lead extends from $L+1<n<\infty $. The purpose of
these leads is to carry the incoming, the reflected and the transmitted
waves. Here $E$ (see Eq.(\ref{tbe1})) is the Fermi energy and without any
loss of generality we choose $\epsilon _n=0$ everywhere in the leads.

In order to calculate the transmission amplitude $t_L$ of a segment
containing $L$ sites we inject a particle from $-\infty $ with an energy $%
\tilde{E}=2cosq$ towards the sample. While the particle passes through the
sample it undergoes multiple elastic scattering. Eventually, it comes out of
the sample from the right end with amplitude $t_L$. Following Pichard \cite
{P86} we write the transmition coefficient $T_L=|t_L|^2$ in terms of the
matrix elements of the total transfer matrix $P_L=\prod_{n=1}^L{\em M}_n$ as 
\begin{equation}
T_L=\frac{4|sinq|^2}{%
|(P_L)_{21}-(P_L)_{12}+(P_L)_{22}e^{iq}-(P_L)_{11}e^{-iq}|^2}.
\label{trans1}
\end{equation}

In the Hamiltonian map approach, the above system corresponds to the
parametric linear oscillator of section 2.2 where the strength $A_n$ of the
instant linear kick (see Eq. (\ref{mmap2a})) is equal to zero for times $%
t\leq 0$ or $t\geq L+1$ describing free rotations in phase plane, while
in the time interval $1\leq t\leq L$ the strength $A_n$ is determined by the
disordered site energy $\epsilon _n$ of the underlying one-dimensional
Schr\"{o}dinger equation (\ref{tbe1}).

In order to establish a relation for the transmition coefficient in the
frame of our Hamiltonian map approach, we recast the two successive maps of
Eqs (\ref{mmap2a}),(\ref{mmap2b}) to the following two-dimensional map $Q_n$ 
\begin{equation}
\left( 
\begin{array}{c}
x_{n+1} \\ 
p_{n+1}
\end{array}
\right) =\left( 
\begin{array}{cc}
cos\mu _0+A_nsin\mu _0 & sin\mu _0 \\ 
A_ncos\mu _0-sin\mu _0 & cos\mu _0
\end{array}
\right) \left( 
\begin{array}{c}
x_n \\ 
p_n
\end{array}
\right) \,,  \label{plom}
\end{equation}
which is related to the transfer matrix $M_n$ defined in Eq.~(\ref{eq:mmap})
through a similarity transformation $R$ 
\begin{equation}
Q_n=RM_nR^{-1};\,\,\,\,R=\left( 
\begin{array}{cc}
1 & 0 \\ 
{\frac{cos\mu _0}{sin\mu _0}} & -{\frac 1{sin\mu _0}}
\end{array}
\right)   \label{simtr}
\end{equation}
From the above Eq.(\ref{simtr}) and Eq.(\ref{trans1}), one obtains for the
transmition coefficient $T_L$ of a system with $L$ scatterers: 
\begin{equation}
T_L=\frac 4{\left( (F_L)_{11}^2+(F_L)_{21}^2\right) +\left(
(F_L)_{12}^2+(F_L)_{22}^2\right) +2}  \label{trans2}
\end{equation}
where the matrix $F_L$ is the product transfer matrix i.e $F_L=\prod_{n=1}^L%
{\em Q}_n$. From Eq.(\ref{trans2}) we see that the sum inside the first
parenthesis in the denominator is equal to the inner product of the vector $%
v(t=L)=F_L\left( 
\begin{array}{c}
1 \\ 
0
\end{array}
\right) $ i.e. to the modulus square of the vector $v(0)=\left( 
\begin{array}{c}
1 \\ 
0
\end{array}
\right) $ evolved under the dynamical map (\ref{plom}) (or equivalently
under the map (\ref{mmap2a}, \ref{mmap2b})) for time $t=L$, in the phase
space of the parametric linear oscillator described by the Hamiltonian (\ref
{ham2}). Similarly, the sum inside the second parenthesis in the denominator
in Eq.(\ref{trans2}) corresponds to the modulus square after the evolution
for time $t=L$ of the initial vector $u(0)=\left( 
\begin{array}{c}
0 \\ 
1
\end{array}
\right) $. It is interesting to note that the initial vectors $v(0),u(0)$
correspond to the unit vectors pointing to the two perpendicular directions
on the phase-space plain.

Using these observations, we can give a geometrical interpretation for Eq.(%
\ref{trans2}) viz. to relate $T_L$ with areas in the phase-space of the two
dimensional Hamiltonian map (\ref{ham2}). In particular, we can interpret
the sum inside each parenthesis in the denominator, as the area of a circle
described by a radius $r_1,r_2$ which is given by the time evolution (under
the map (\ref{plom}) ) of the initial vectors $(r_{1,2},\theta
_{1,2})_{t=0}^{}=(1,0),(1,\pi /2)$. Thus Eq.(\ref{trans2}) can be rewritten
in the following form 
\begin{equation}
T_L={\frac{4\pi }{\pi r_1^2+\pi r_2^2+2\pi }}={\frac{2S_{tot}^0}{%
S_{tot}^0+S_{tot}^L}}  \label{trans3}
\end{equation}
where $S_{tot}^L=S_1^L+S_2^L$ is the sum of the areas defined by the radius $%
r_1,r_2$ at time $t=L$. In the case of perfect lattice where we have simple
rotations of the initial vectors $(r_{1,2},\theta _{1,2})^0$, the areas
defined after time $L$ will be the same $S_{tot}^L=S_{tot}^0=2\pi $ and
hence $T_L=1$.

Using Eq. (\ref{rmap}) we can write Eq.(\ref{trans3}) in a way that it is
more tractable for numerical calculations, i.e. 
\begin{equation}
T_L={\frac 4{\prod_{n=0}^{L-1}\left( D_n^{(1)}\right)
^2+\prod_{n=0}^{L-1}\left( D_n^{(2)}\right) ^2+2}}  \label{tnum}
\end{equation}
where $D^{i=1,2}$ correspond to initial conditions $(r_i,\theta
_i)^0=(1,0),(1,\pi /2)$ respectively.

The results we have obtained so far provide an exact, although non-closed,
analytical description of any one-dimensional system that can be written in
the tight-binding form of Eq.(\ref{tbe2}). We will now evaluate them for the
specific case of RDKP to describe those relevant features of the
transmission coefficient that may be the fingerprint of extended states.

In Fig.~6a,b we show numerical results for a system of 10000 scatterers
after averaging over more than 10000 different realizations of the
disordered sample. We take the values of $\epsilon _1=0$ and $\epsilon
_2=9.75885$ and the defect concentration $Q=0.5,$ i.e. the most random case.
In Fig.~6a we used the critical energy $q_{cr}=5.443223$ lying inside the
second zone of the spectrum while Fig.~6b $q_{cr}=33.13$ lying inside the $%
10^{th}$ zone. We note that states close to the resonant energies have very
good transmission properties, similar to those at the resonant energy where
the transmition coefficient $T$ is equal to one. This is compatible with the
findings of the random dimer model with one-band (tight-binding
approximation) \cite{find}. Moreover in RDKP model, the width of the peaks
depends on the order of the resonance, as mentioned previously (see also 
\cite{SMA94}). From the comparison between Fig.~6a and Fig.~6b we see that
the higher the resonance the wider the band of states with $T\approx 1$. In
Fig. 6c we presented results for a different defect concentration $Q$ in
order to study the dependence of the transmition coefficient depends on $Q$.
We use the same values of $\epsilon _1,\epsilon _2$ and $Q=0.2$. By
comparison with Fig.~6a we conclude that as $Q$ decreases the number of
transparent states, i.e states with transmition coeffissient close to one $%
T\sim 1$, increases, in perfect agreement with the results of the previous
section for the localization length of nearly-resonant states.


\section{Conclusions}


We have studied a Kronig-Penney model with binary on-site disorder randomly
assigned every two sites. For such a model it was found \cite{SMA94} that
there exist infinite number of special energies $E_{cr}=q_{cr}^2$ at which
transparent states appear. We recover these results using a new approach
based on classical Hamiltonian maps. We generalized our results for a $N$%
-mer case and obtained a simple expression for the resonant energy values.
We constructed a phase diagram in the $\epsilon _{cr}-N$ plane which marks
three distinct regions: one where no resonant states appear, the second
that contains some of the resonant states and the third that contains all
possible resonant states for different correlated blocks of size $N$. This
separation in three distinct regions is valid only in the first zone and as
a result it might have some relevance to the low temperature system
properties. Our dynamical system approach maps resonant delocalized states
to bounded trajectories, while localized states are represented by unbounded
trajectories in the phase space $(p,x)$. Making use of an expansion in the
vicinity of the resonance, we derived an analytical expression of Eq. (\ref
{estimate}) for the Lyapunov exponent for the nearly resonant states.
Finally, in the frame of our Hamiltonian map we established a simple
geometrical picture for the transmition coefficient showing that it
corresponds to evolution of areas in the phase space of a linear parametric
oscillator. Using these last results we calculated the transmission
coefficient that exhibits peaks up to $T=1$ for energy values equal to the
resonant ones. Near the resonant energies there are nearly-transparent
states with large transmition coefficient, the number of which is inversely
proportional to the defect concentration $Q$ and increases with the
resonance order. The properties of the RDKP model that were analyzed in this
work could be used to several mesoscopic quasi-one dimensional studies.


\subsection{Acknowledgment}


We acknowledge support of $\Pi$ENE$\Delta$ 95-115 grant of the General 
Secretariat for Research and Technology of Greece.
One of the authors (F.\thinspace M. \thinspace I.) wishes to acknowledge
support of Grant ERBCHRXCT\thinspace 930331 Human Capital and Mobility
Network of the European Community; and the support of Grant No RB7000 from
the International Science Foundation. (T.\thinspace K.) acknowledges with
thanks many useful discussions on the Anderson localization with A. Politi
during his visits to the Istituto Nazionale di Ottica. 

\newpage
\centerline{\bf Figure Captions} \vspace{1.0cm}

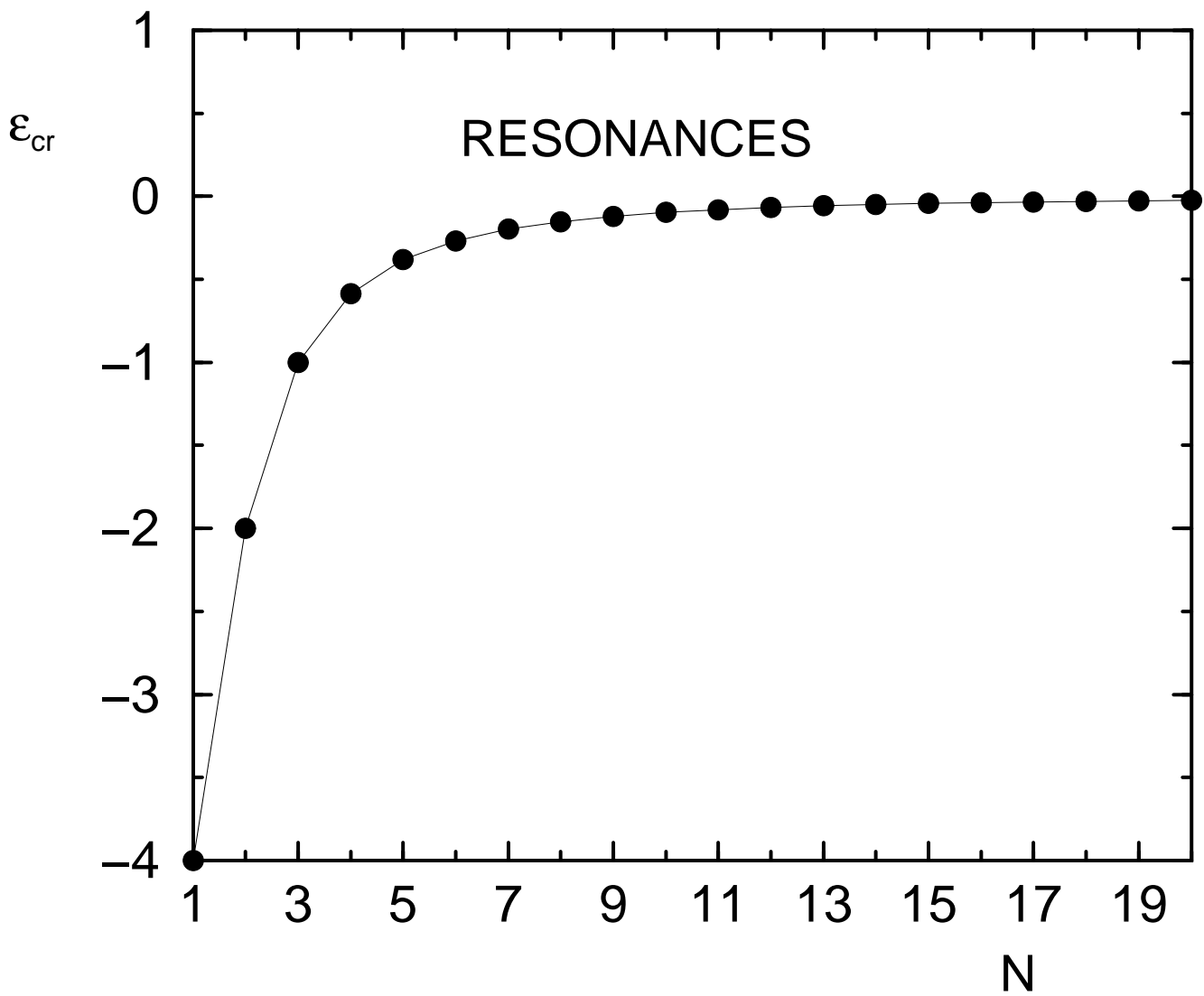
\noindent {\bf Figure 1:} The phase diagram showing the critical $\epsilon
_{cr}$ values as a function of the block size $N$.\\ 

\noindent {\bf Figure 2:} Phase space of the map of Eq. (13) for $p_0=x_0=1$:%
\\ 

(a) One value of $\epsilon _2$ in the sequence $\epsilon :$...$\epsilon
_1\epsilon _1\epsilon _1\epsilon _2\epsilon _1\epsilon _1\epsilon _1...$ for 
$q_{cr}=11.4;\,$

$\epsilon _1=0;\epsilon _2=9.75885$ ; We note that one point is outside of
the ellipse representing the kick to the $\epsilon_1$ trajectory by $%
\epsilon_2$.\\

(b) Two values of $\epsilon _2$ (one dimer) in the sequence $\epsilon
:...\epsilon _1\epsilon _1\epsilon _1\epsilon _2\epsilon _2\epsilon
_1\epsilon _1\epsilon _1...$ for $\epsilon _1=0$ ; $\epsilon _2=9.75885$; $%
q_{cr}=11.4$. We note that there is a point inside the ellipse representing
the kick to the $\epsilon _1$ trajectory by the first $\epsilon _2$ value.
The second $\epsilon _2$ value kicks the trajectory back to the ellipse.\\ 

(c) Dimers of type $\epsilon _2$ , randomly (with probability $Q=0.5)$
distributed in the sequence $\epsilon $ for $\epsilon _1=0$, $\epsilon
_2=9.75885$ and $q_{cr}=11.4$.\\ 

\noindent {\bf Figure 3:} The phase space of the map (16-17) for $p_0=x_0=1$
and $Q=0.5,\epsilon _1=0$ :\\ 

(a)$N=2$ ( dimer); $q_{cr}=11.4$; $\epsilon _2=9.77885;$\\

\thinspace (b) $N=3$ (trimer); $q_{cr}=1.489$; $\epsilon _2=1.25;$\\

\thinspace The length of sequence $\epsilon $ is equal to $L=1000$ .\\ 

\noindent {\bf Figure 4:} Nearly resonant states for a dimer $(N=2)$.
Comparing with the resonant states shown in Figs.2-3a, nearly-resonant
states correspond to the unbounded (for $t \rightarrow \infty $) motion with
a slow spread of the points in the phase space:\\

\thinspace (a) $q=11.399$ ; $\epsilon _2=9.75885;$\\

(b) $q=11.390$ ; $\epsilon _2=9.75885;$\\

(c) $q=11.38$ ; $\epsilon _2=9.75885;$\\

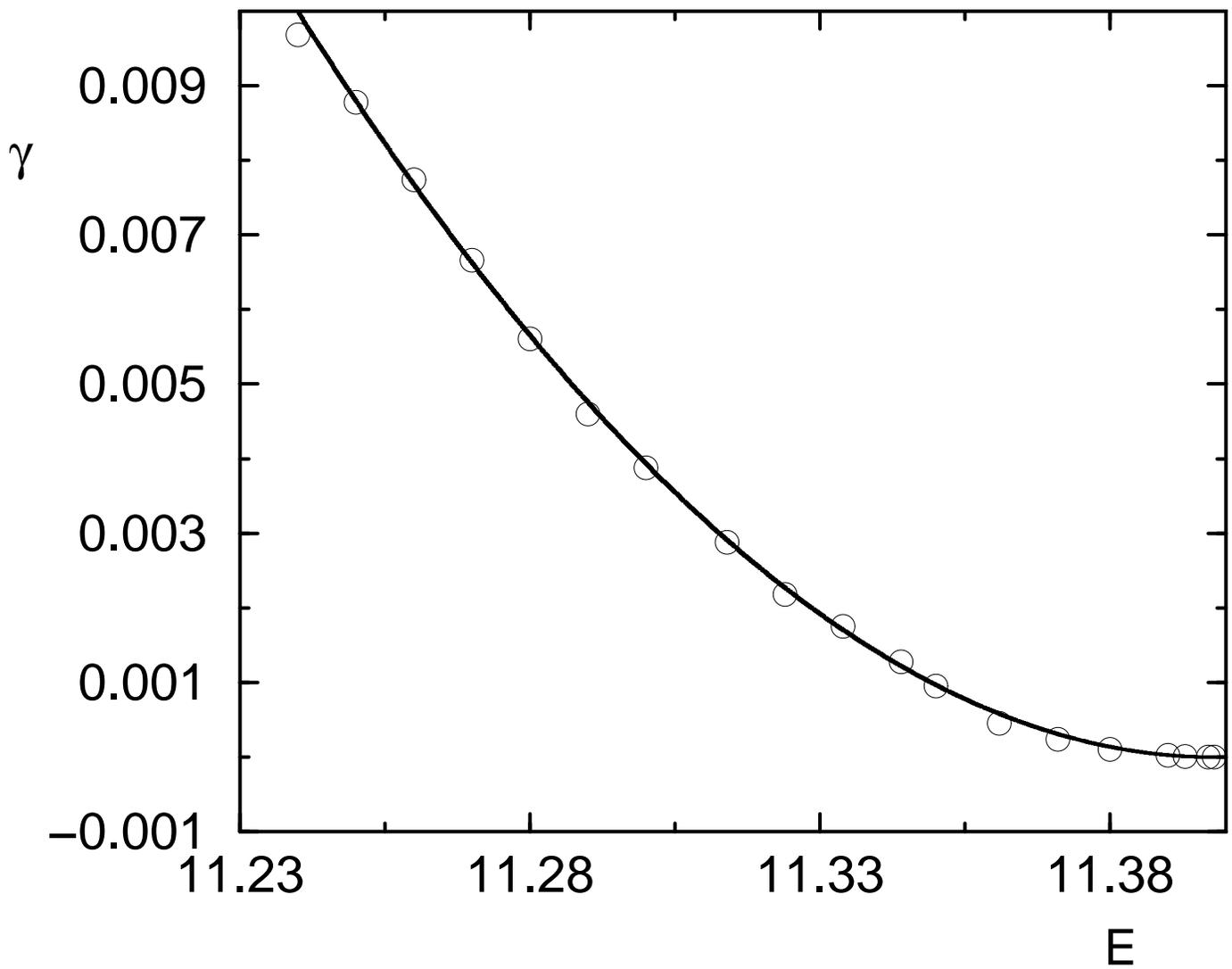
\noindent {\bf Figure 5:} Numerical (circles) and analytical estimation
(solid line) of the Lyapunov exp. for near resonant states. We use $%
q_{cr}=11.4$. We observe good agreement.\\ 

\noindent {\bf Figure 6:} Transmission coefficient (averaged over more than
10000 realizations) as a function of $q=E^2$ for a system with $\epsilon _1=0
$, $\epsilon _2=9.75885$:\\ 

\thinspace (a)$Q=0.5$ and $q_{cr}=5.44$ corresponding to a resonance inside
the second band.\\ 

(b)$Q=0.5$ and $q_{cr}=33.13$ corresponding to a resonance inside the $%
10^{th}$ energy band. We note that with respect to (a) the band of states
with $T\approx 1$ is wider.\\ 

(c)$Q=0.2$ and $q_{cr}=5.44$ corresponding to a resonance inside the second
energy band. We note that with respect to (a) we have now more states with $%
T\approx 1$. \\ 


\newpage

\begin{thebibliography}{99}

\bibitem{Dunlap}  D. Dunlap, H.-L. Wu and P. Phillips, Phys. Rev. Let. {\bf %
65}, 88 (1990). 

\bibitem{Phillips}  P. Phillips and H.-L. Wu, Science {\bf 252}, 1805
(1991). 

\bibitem{Bovier}  A. Bovier, J. Phys. A {\bf 25}, 1021 (1992). 

\bibitem{flor}  J. C. Flores, J. Phys. Cond. Matt. {\bf 1}, 8471 (1989). 

\bibitem{Gangopadhyay}  S. Gangopadhyay and A. K. Sen, J. Phys. Cond. Matt. 
{\bf 4}, 9939 (1992). 

\bibitem{Evangelou1}  S. N. Evangelou and E. N. Economou, J. Phys. A {\bf 26}
, 2803 (1993). 

\bibitem{Evangelou2}  S. N. Evangelou and A. Z. Wang, Phys. Rev. B {\bf 47 }
, 13126 (1993). 

\bibitem{find}  (a) P. K. Datta, D. Giri and K. Kundu, Phys. Rev. B {\bf 47}%
, 10727 (1993); (b) P. K. Datta, D. Giri and K. Kundu, Phys. Rev. B {\bf 48}%
, 16347 (1993) 

\bibitem{IKT95}  F.M.Izrailev, Tsampikos Kottos and G. P. Tsironis, Phys.
Rev. B {\bf 52}, 3274 (1995); J. Phys.: Condens. Matter {\bf 8}, 2823
(1996). 

\bibitem{SMA94}  Angel S\'{a}nchez, Enrique Maci\'{a} and Francisco
Dom\'{i}nguez-Adame, Phys. Rev. B, {\bf 49}, 147 (1994). 

\bibitem{BFLT82}  J. Bellissard, A. Formoso, R. Lima and D. Testard, Phys.
Rev. B {\bf 26}, 3024 (1982). 

\bibitem{P86}  J. L. Pichard, J. Phys. C : Solid State Phys. {\bf 19}, 1519
(1986). 

\end{thebibliography}
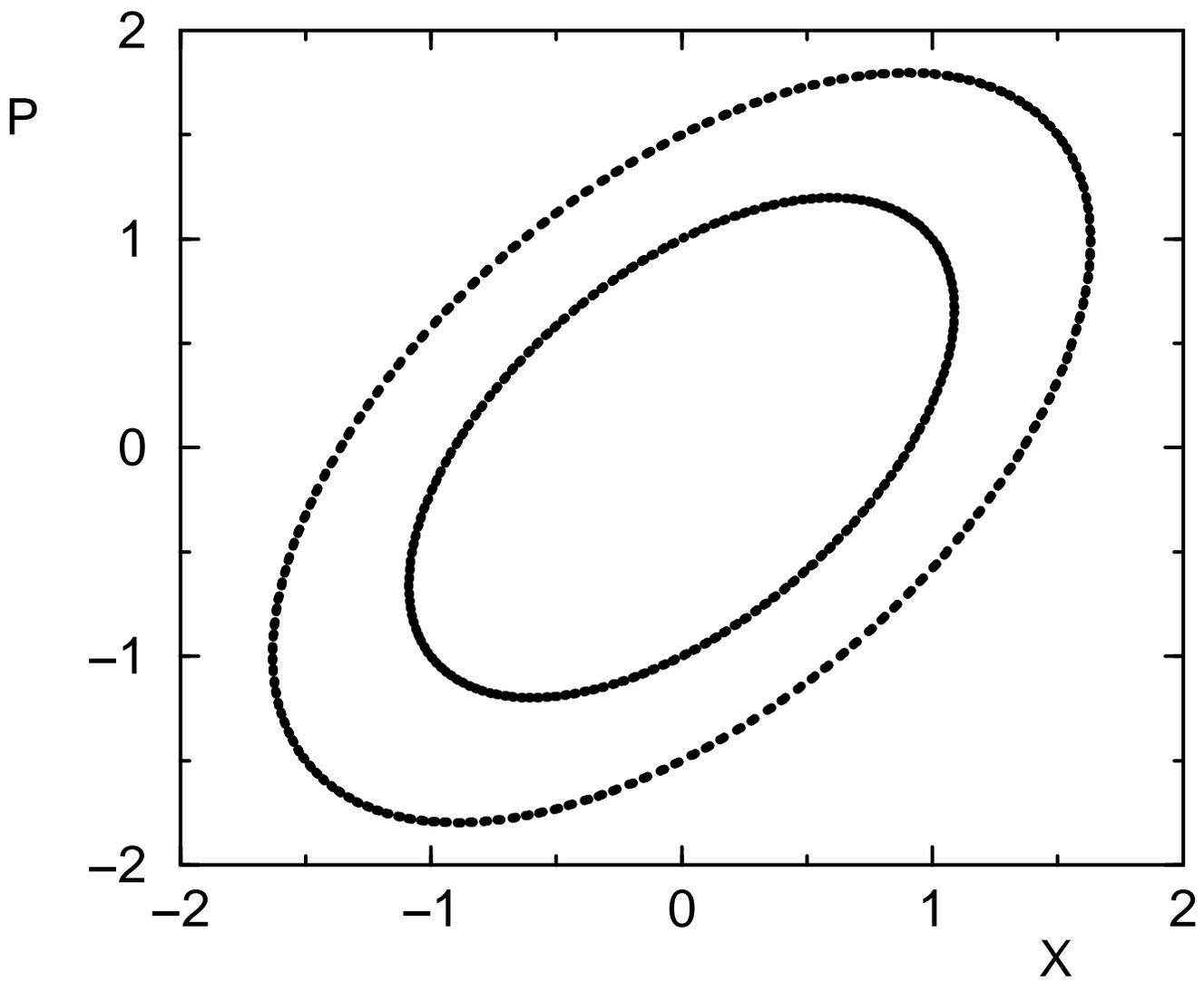
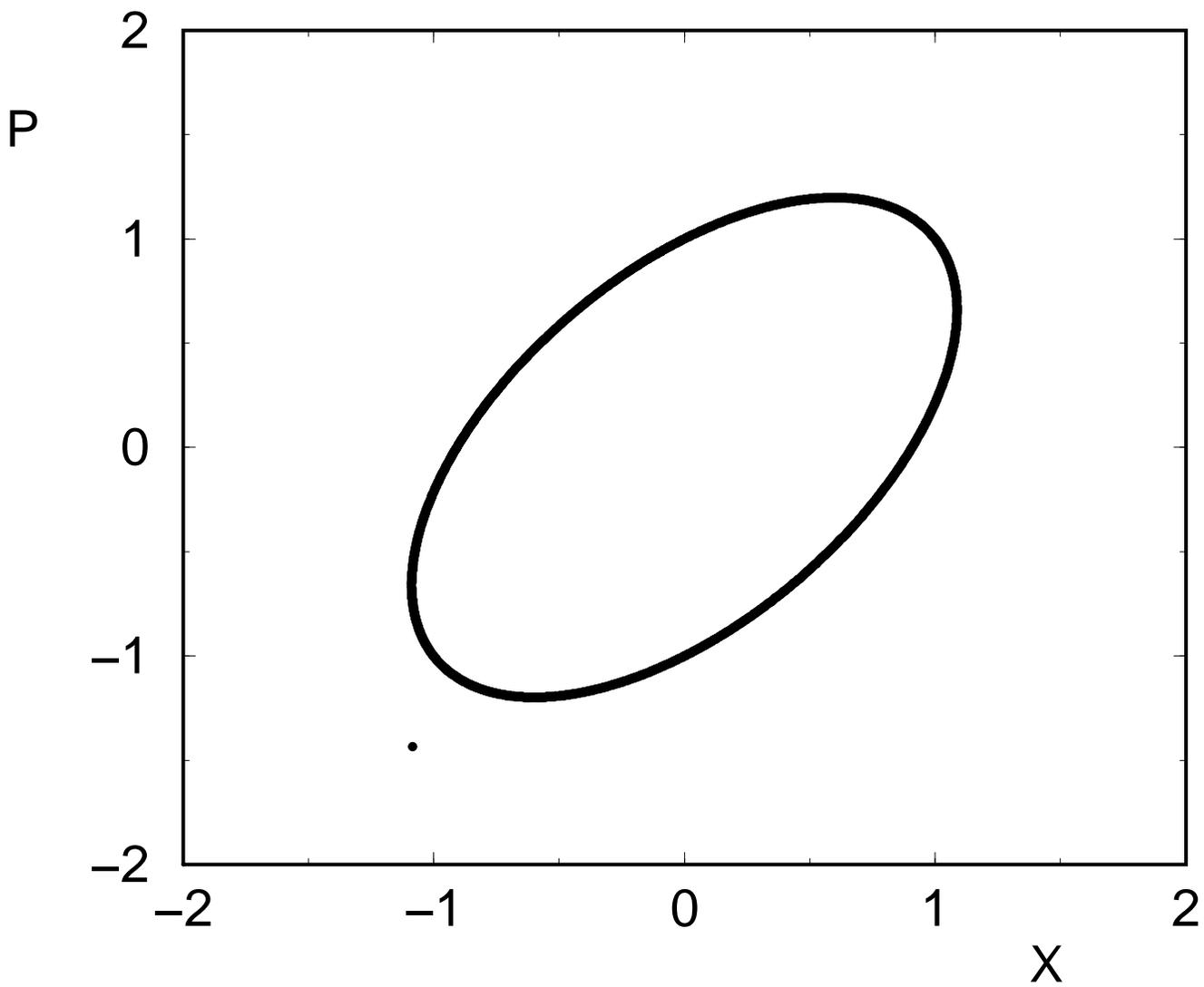
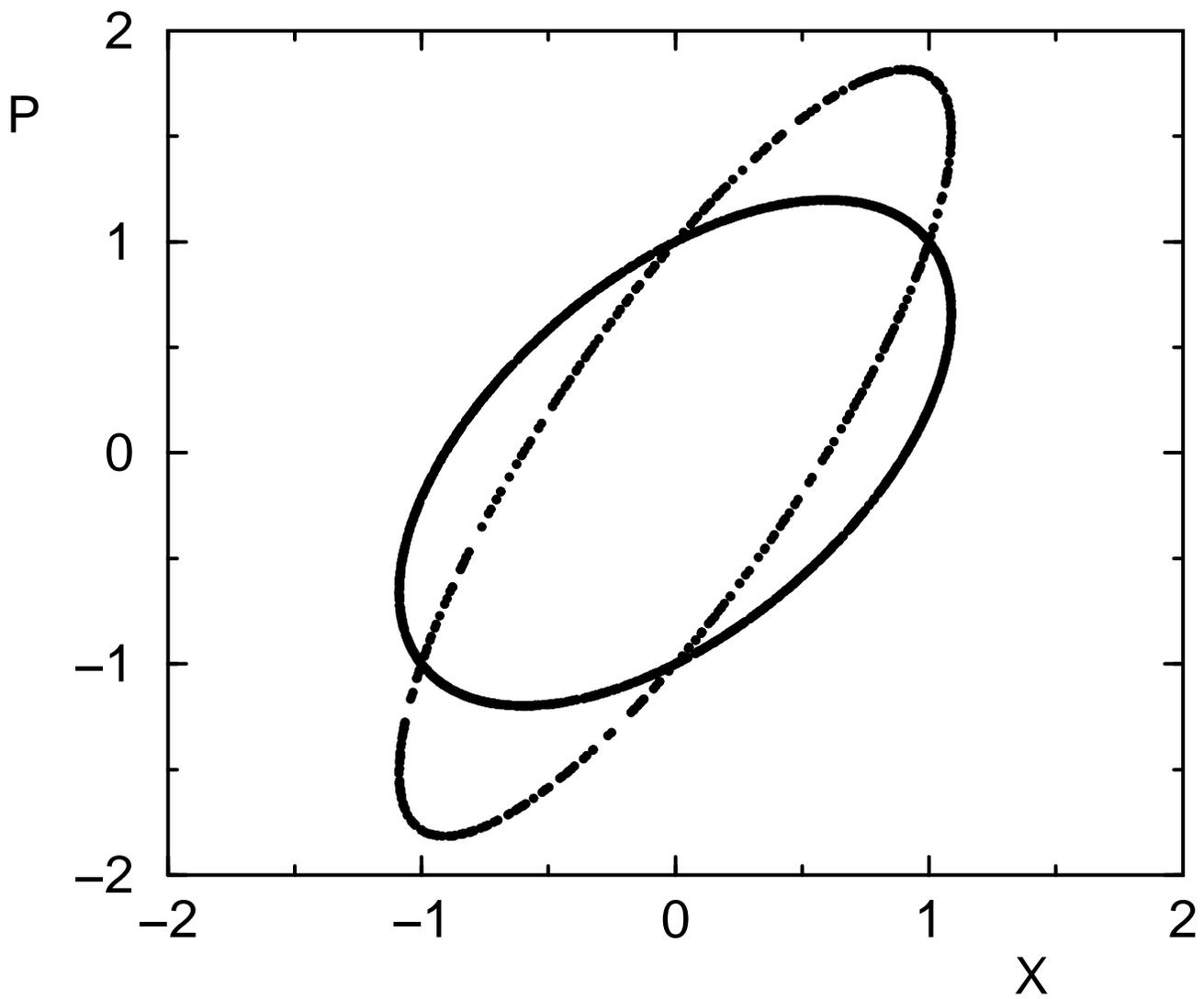
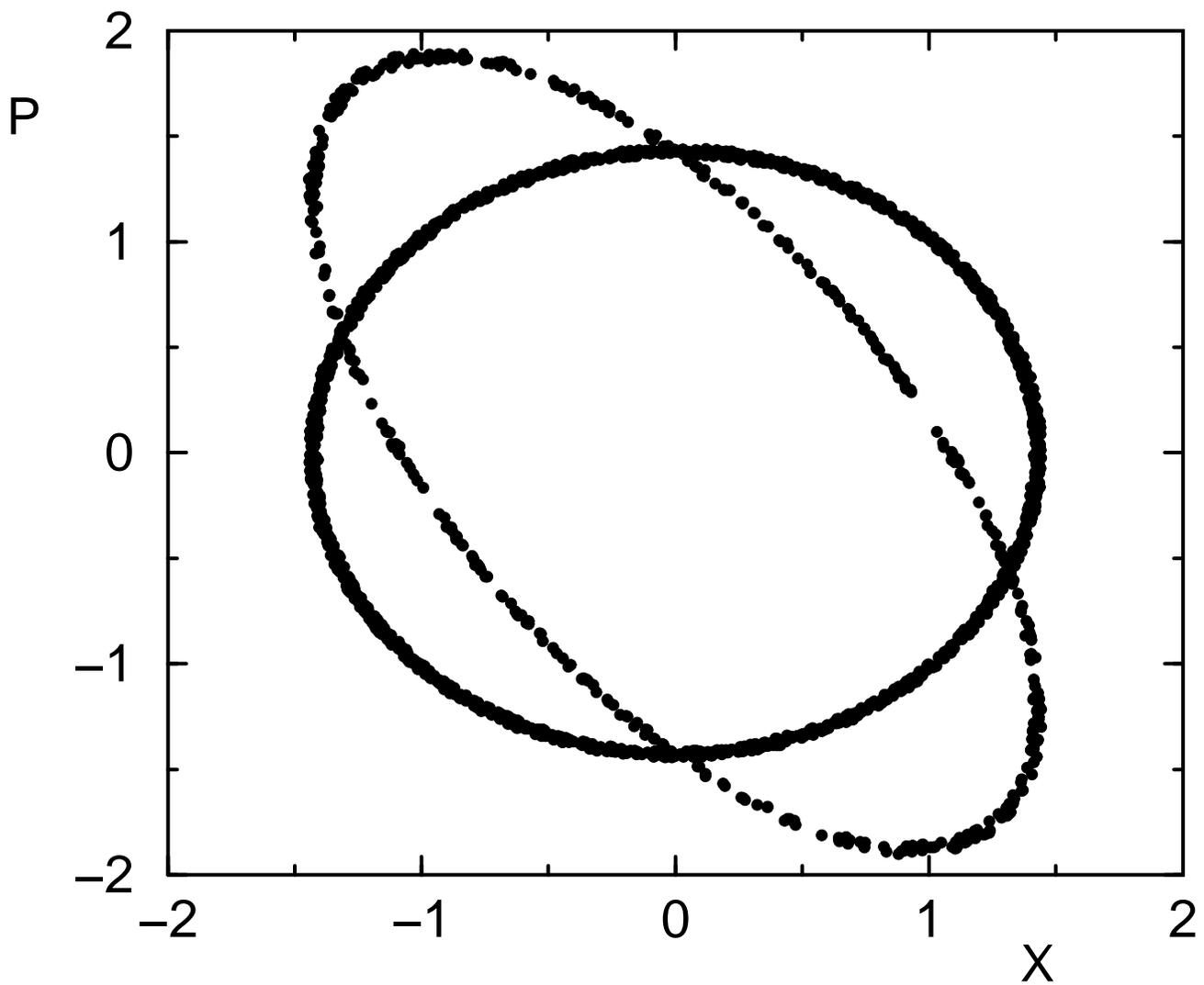
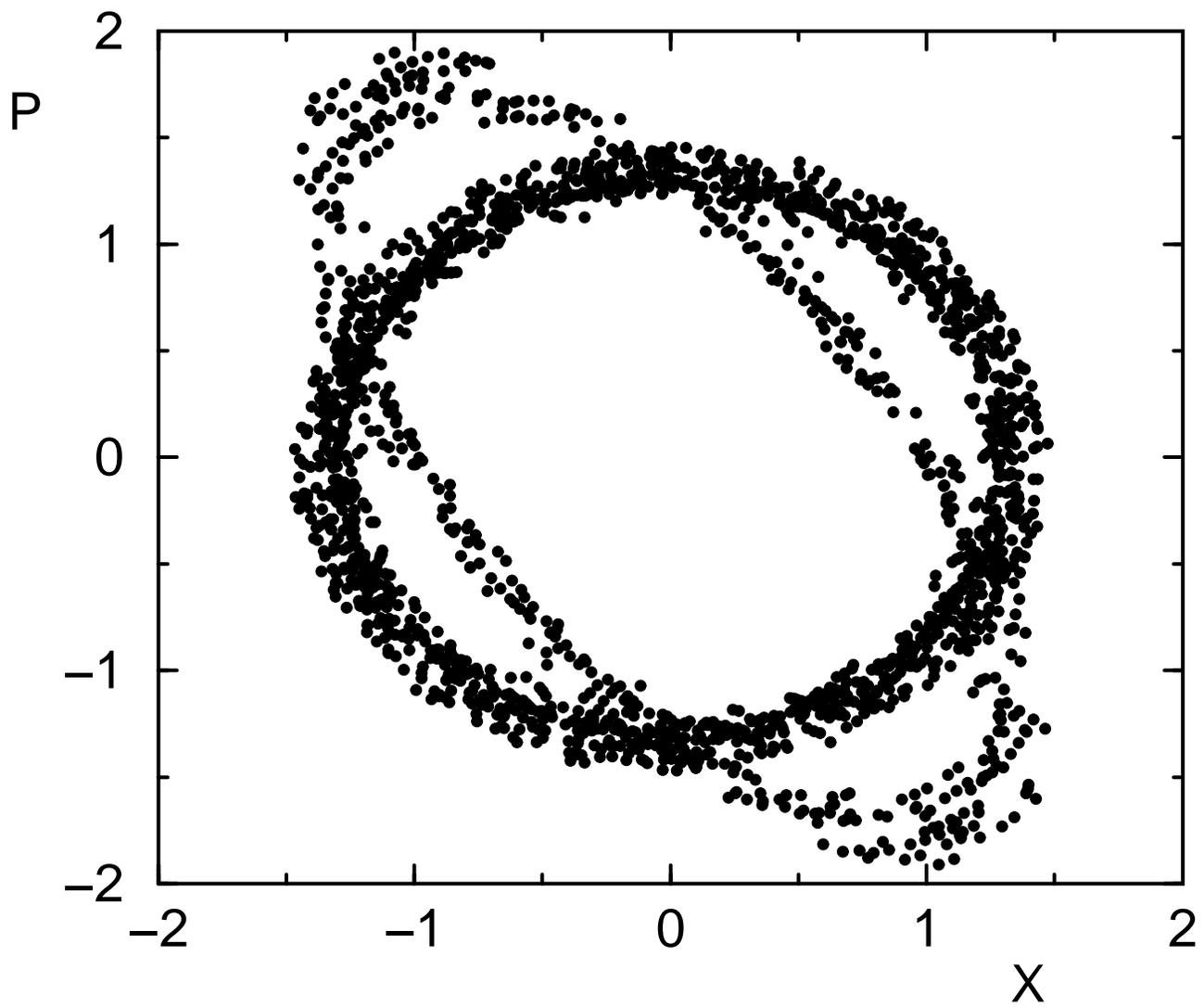
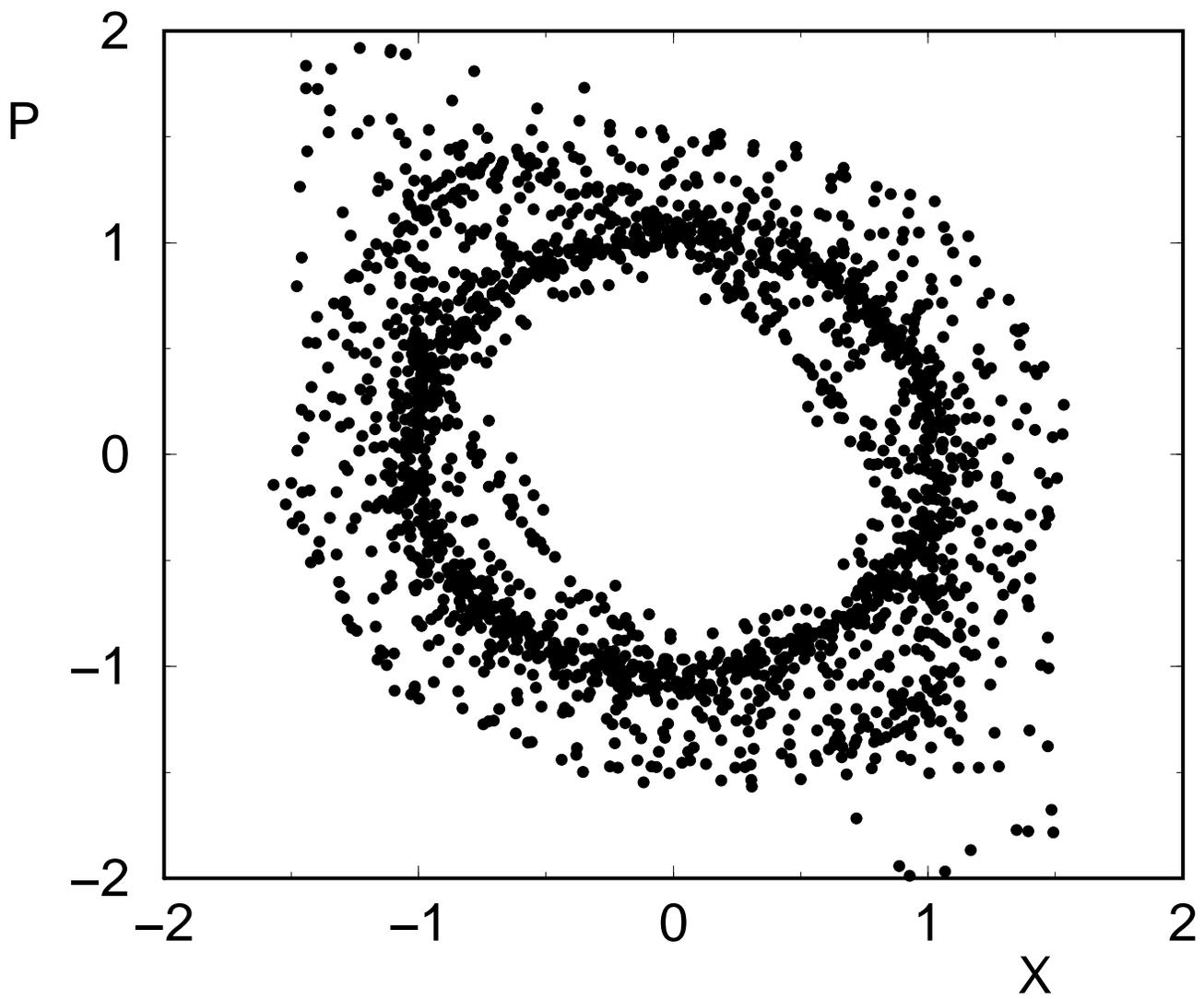
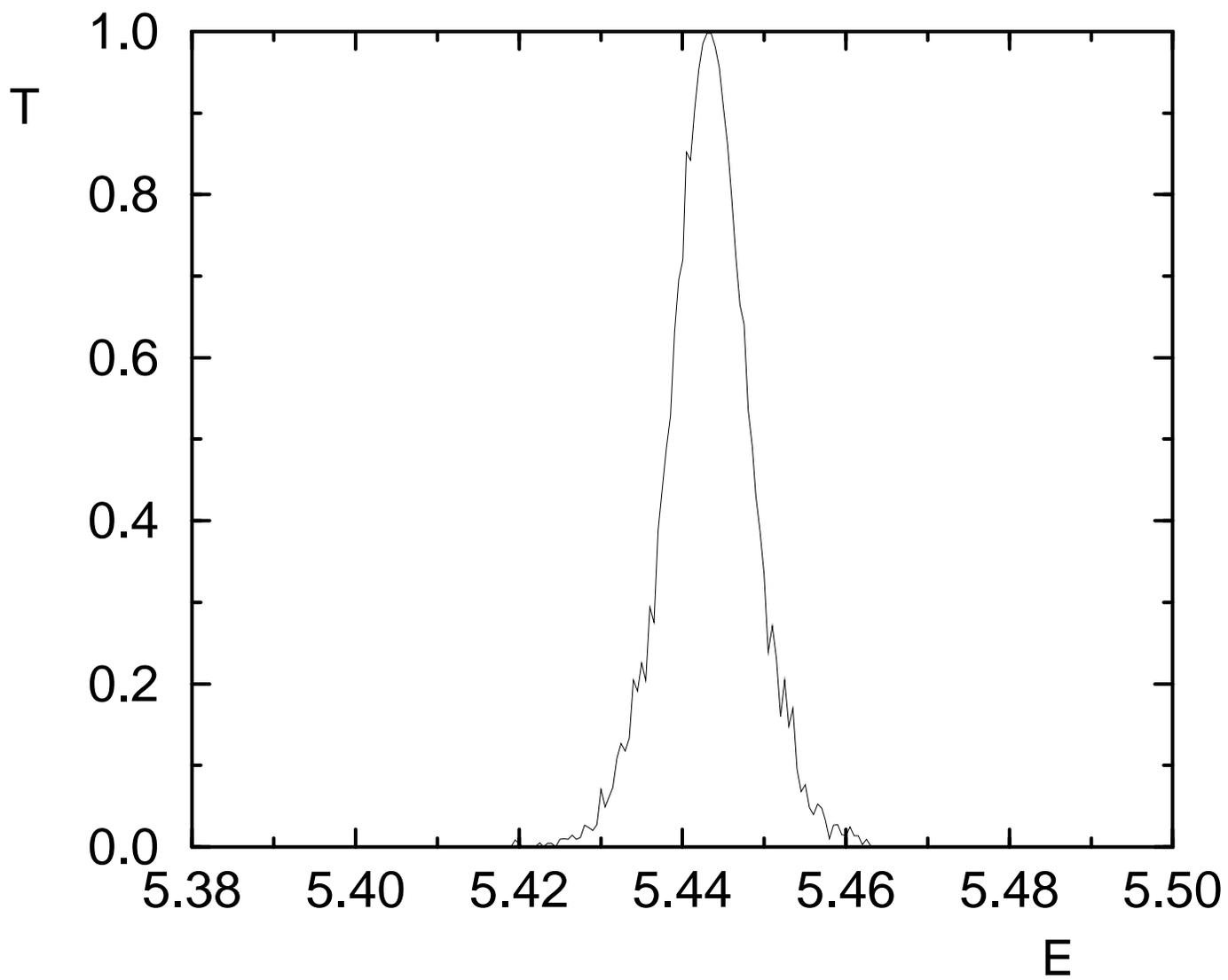
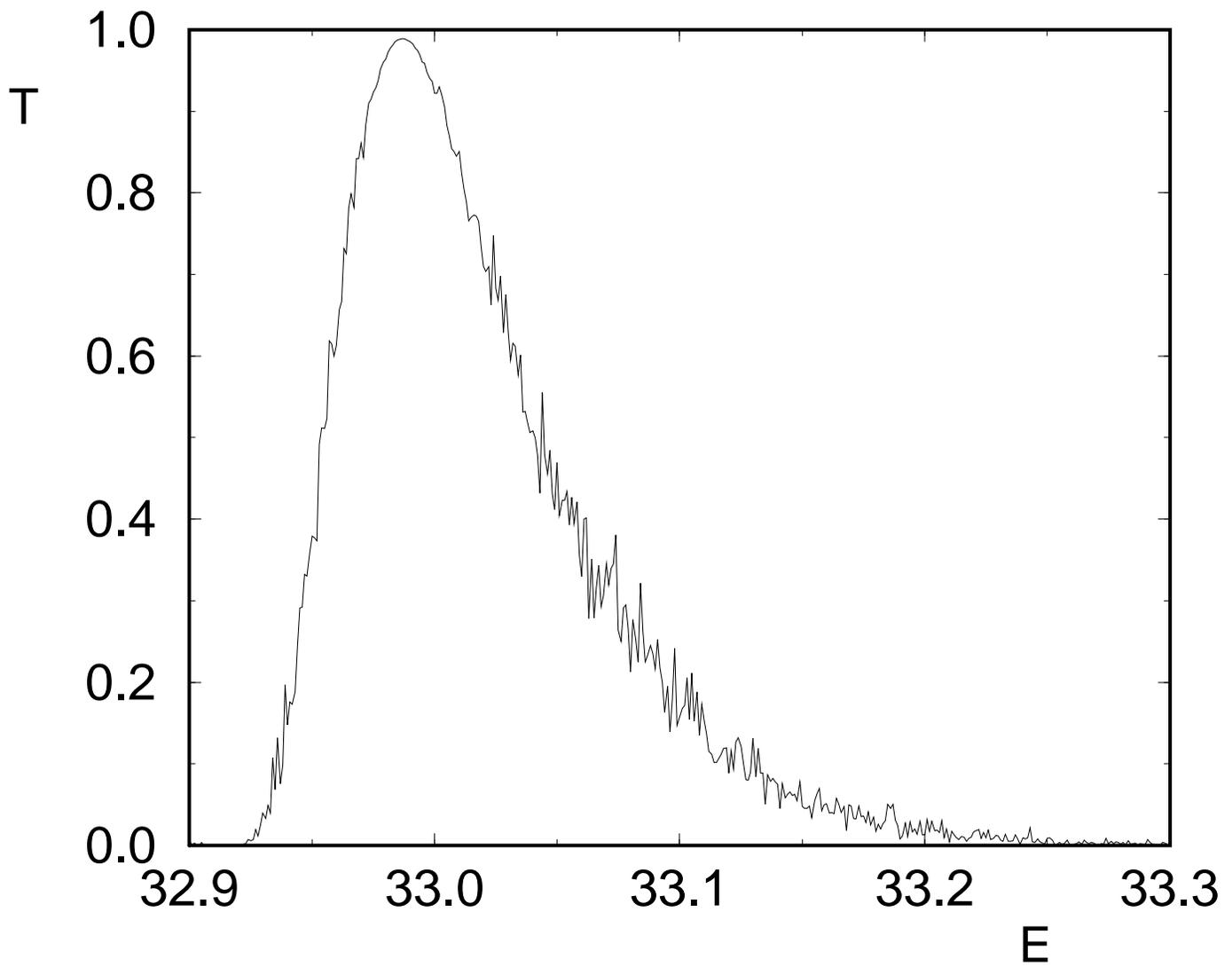
\end{document}